\documentclass[12pt,preprint]{aastex}

\shorttitle{The mass function of M22}
\shortauthors{Albrow et al.}

\begin{document}

\title{The Spatially-Resolved Mass Function of the Globular 
Cluster M22\altaffilmark{1}}

\altaffiltext{1}{Based on observations with the NASA/ESA {\it Hubble Space Telescope} 
                 obtained at ST ScI, which is operated by AURA, Inc. under 
                 NASA contract NAS 5-26555.}

\author{Michael D. Albrow\altaffilmark{2,3},
  Guido De Marchi\altaffilmark{3}, Kailash C. Sahu\altaffilmark{3}} 

\altaffiltext{2}{Department of Physics and Astronomy, University of Canterbury,
                 Private Bag 4800, Christchurch, New Zealand}

\altaffiltext{3}{Space Telescope Science Institute,
                 3700 San Martin Drive, Baltimore, MD 21218}

\email{m.albrow@phys.canterbury.ac.nz, demarchi@stsci.edu, ksahu@stsci.edu}

\begin{abstract}

HST imaging of M22 has allowed, for the 
first time, a detailed and uniform mapping of mass segregation in a
globular cluster. Luminosity and mass functions 
from the turnoff down to the mid to lower main sequence are
presented for M22 in annular bins from the centre of the cluster
out to five core radii. Within the core, a significant enhancement 
is seen in the proportion of 0.5-0.8 $M_{\sun}$ stars compared
with their numbers outside the core. Numerical modelling of the
spatial mass spectrum of M22 shows that the observed degree of
mass segregation can be accounted for by relaxation processes
within the cluster. The global cluster mass function for M22
is flatter than the Salpeter IMF and cannot be represented by a single power 
law.

\end{abstract}


\section{Introduction}

As in many areas of astronomy, the advent of the Hubble Space Telescope
has revolutionised the study of globular clusters. Primarily because of
crowding, ground-based observations of the central regions of globular
clusters are limited to brighter stars, at or
above the main sequence turnoff. HST allows access to the study of
stellar populations below the turnoff including main sequence stars and
white dwarfs.

Main-sequence stars below the turnoff in globular clusters (typically
$m < 0.8 M_{\sun}$) have evolved little from their initial zero-age
main-sequence (ZAMS) state. Thus, mass functions derived from globular cluster 
luminosity functions can be used as indicators
of a stellar initial mass function (IMF).
Most notably in recent years, several groups have used HST WFPC2
photometry to probe mass and luminosity functions for several
globular clusters down to the hydrogen burning limit. For example,
\citet{Paresce2000} have documented the turnover in the
luminosity function at $\sim 0.3 M_{\sun}$ for a sample of twelve
Galactic globular clusters. 
In NGC 6397 \citet{King1998} found that the mass function
increases slowly for masses down to 0.1 $M_{\sun}$ and then drops 
rapidly.

Although individual globular cluster main sequence stars are
little evolved from the ZAMS, the main sequence itself has been subject 
to modification by cluster dynamical effects. These include not only
intra-cluster effects such as relaxation due to two-body 
interactions but also tidal interactions between a globular
cluster and its Galactic environment. 
Relaxation of globular clusters has been studied in detail 
through dynamical equilibrium models \citep{King1966,Gunn1979}
and through direct
numerical n-body simulations \citep{Aarseth1999}.
A comprehensive review of globular cluster dynamics is given by \citet{Meylan1997}.
Briefly, two-body interactions tend to transfer kinetic energy outward
from the core and produce mass segregation, a depletion of the relative
fraction of low
mass stars in the central regions relative to their proportions
outside the core. 
Only since the mid-1990's has this effect been reliably
observed in globular cluster cores, for example in 47 Tuc
\citep{Paresce1995}, NGC 6752 \citep{Shara1995} and NGC 6397
\citep{King1995}.
(Note that the core of a globular cluster is usually parameterised by
the core radius, $r_{c}$, defined by \citet{King1962} as the scale factor in his
empirical formula for the surface density profile.) 
The most important external dynamic effect is
disk shocking, which tends to strip the lightest stars out of a
globular cluster during orbital crossings of the Galactic plane.
To best avoid both internal and external dynamical modifications,
the stellar luminosity functions in globular clusters should be
obtained at radii close to
the half-light radius of the cluster \citep{Lee1991,Paresce2000}.

A further complication in deriving a global IMF is the
presence of binary main-sequence stars in a globular cluster. 
Near-equal-mass binary stars appear on a color-magnitude diagram 
in a main sequence displaced upwards by 0.75 mag \citep{Elson1998}.
In only a few cases, for example NGC 6752 \citep{Rubenstein1997}, has 
the photometry been sufficiently precise to resolve this binary
main sequence. Normally, the presence of binary stars will
contaminate a main-sequence luminosity function, particularly
in the core of a cluster where, due to mass segregation effects,
the binary fraction is highest. 
In 47 Tuc, \citet{Albrow2001}
found the fraction of binary stars to be around 13\% in the
innermost 4 $r_{c}$, with some evidence that this fraction
was highest ($\sim 20\%$) within 1 $r_{c}$, dropping to $\sim 8\%$
at 2.5 $r_{c}$. 
Such a dropoff was also noted by \citet{Rubenstein1997}
in NGC 6752. For globular clusters showing at least a moderate degree
of central concentration, $log(r_{tidal}/r_{c}) \gtrsim 1.5$, 
the half-light radius is 
generally at least several times $r_{c}$ so luminosity functions 
derived at the half-light radius should be reasonably free from binary 
contamination.

In this paper we derive the luminosity and mass functions for M22 (NGC 6656), a
globular cluster located about one third of the way between the Sun
and the Galactic bulge. Our observations (taken as part of another
program) are not particularly deep but cover a large spatial area
from the center out to several $r_{c}$. Our focus is thus on
determining the degree of
mass segregation in the middle to upper main sequence rather 
than on probing the lowest mass stars.
From four fields that we
subdivide into concentric annular radial bins, we determine
how the luminosity and mass functions change with radius in this
cluster. Sections 2 and 3 discuss the data and their reduction. In section
4 and 5 we consider the derivation of the luminosity and mass functions.
In section 6 we compare these results with a dynamical model for the cluster.

\section{\label{sect:Observations}Observations}

As part of a program to detect gravitational microlensing events by
stars within M22 \citep{Sahu2001}, observations were taken during  22 February to 15
June, 1999, using the WFPC2 camera aboard HST.  The images were taken
at 43 epochs, with a typical separation of about 3 days. A subset of 9
images were taken with a separation of about 1 day, which were dithered
at a sub-pixel level.  One additional epoch of observations was taken a
year later, on 18 February 2000. At each epoch, images were taken of
three fields (hereafter referred to as pointings 1-3) in the central
region of 
M22. Most of the observations
were taken in the I (F814W) filter, with every fourth observation in
the wide-V (F606W) filter.  To optimize the overhead and exposure times
during a single orbit, the 3 observed fields were so chosen that  they
used the same guide stars. This avoided the overheads involved in
switching between guide stars during an orbit, but led to slight
overlap between different fields. The orientation of the images was
kept fixed in all the observations. To facilitate cosmic ray removal,
the images were taken in pairs for each filter, each with an
integration time of 260 sec. For each observed field, the total
exposure time is 17160 sec in the F814W filter and 5200 sec in the
F606W filter.

\begin{figure}
\epsscale{0.7}
\plotone{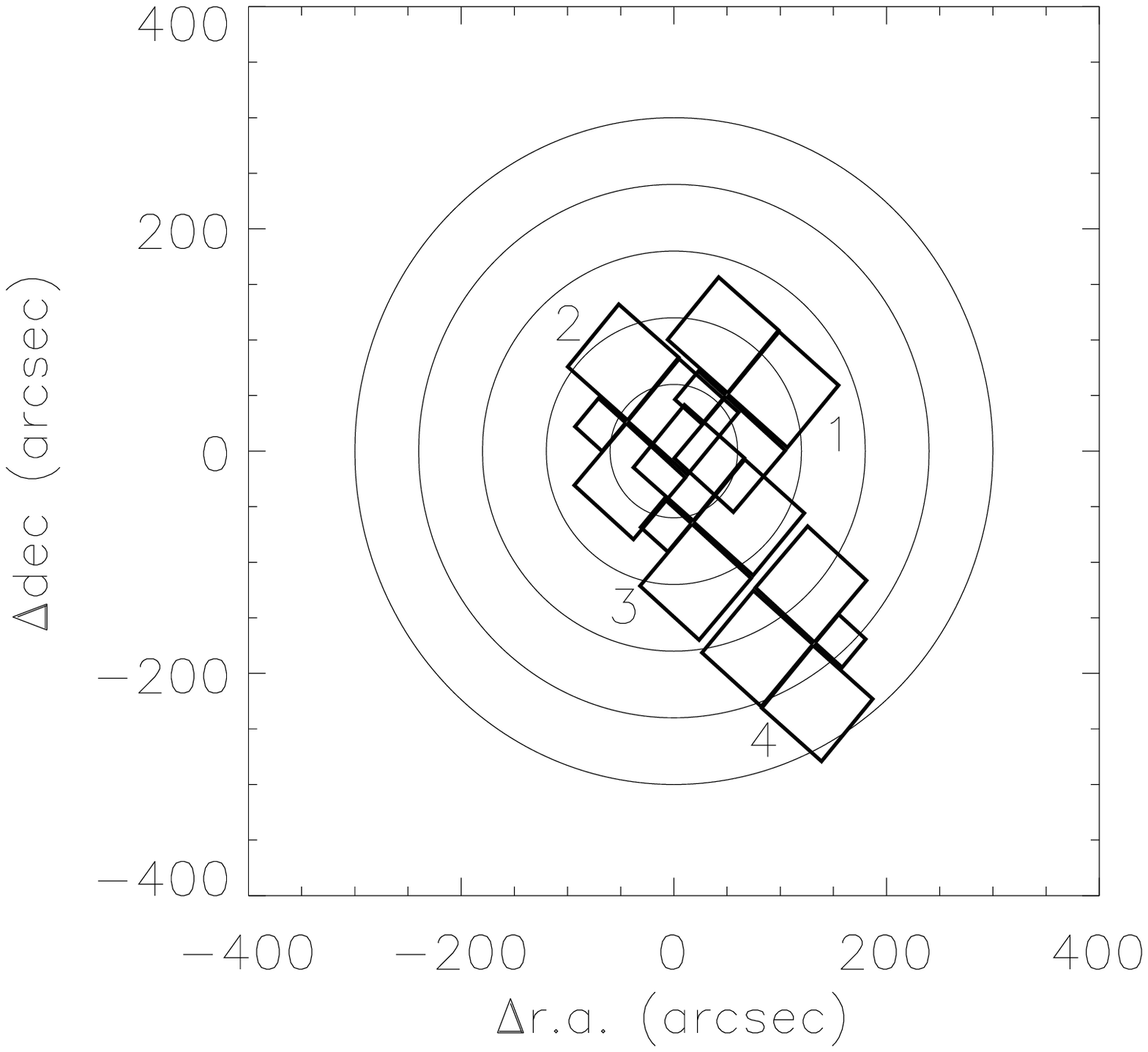}
\caption{\label{fig:m22area}
Area coverage of the 4 WFPC2 pointings relative to 60'' annular bins
around the cluster center.}
\end{figure}

The above observations of the central regions of M22 were supplemented
with exposures from the HST archive of a field (hereafter pointing 4)
at the approximate
half-light radius of the cluster, 3.5' southeast of the cluster center.
These consisted of 4$\times$1200~s exposures in F814W and
2$\times$1100~s + 2$\times$1200~s
exposures in F606W.  A luminosity function from these datasets was derived by
\citet{DeMarchi1997} and later confirmed by \citet{Piotto1999}.
We thus have 16 different pointing/CCD combinations listed in 
Table~\ref{table:CCDfields}.
The four WFPC2 pointings used for this paper 
are shown in Fig.~\ref{fig:m22area} relative to the cluster center which
we take to be at J2000 coordinates (18h36m24.2s,-23$\degr$54'12'') from
\citet{Harris1996}.

Additionally, the HST archival dataset u27xjd01t was used to establish
the luminosity function of the Galactic bulge local to M22. This
archive consists of a single (non-CRSPLIT) 2400~s F814W exposure,
offset from the center of M22 by approximately 9 arcmin to the southwest. 

\section{\label{sect:DataReduction}Data Reduction}

The data frames were initially put through the standard HST on-the-fly
calibration pipeline which involves bias and dark subtraction and
flat-field correction. The remaining steps in the photometric
reduction process were done using the HSTPHOT 1.0 package
\citep{Dolphin2000a}. Data quality images were used to mask 
bad pixels and vignetted regions.  Pairs of images (CR-SPLITs) taken
during a single orbit and with the same dither offset and filter were
combined for cosmic ray removal. Sky images were then calculated and
hot pixels removed.

PSF-fitting photometry was done using the MULTIPHOT task in
HSTPHOT 1.0. This program uses the combined signal from all the images at
a given pointing for object detection. We used a detection threshold
of 3.0 for the minimum signal-to-noise in the combined images.  This
threshold was deliberately set lower than what would eventually be
used in the selection of stars for further analysis in order to
prevent marginally-detected stars from contaminating the measurements
of their neighbours.

The artificial star routine in MULTIPHOT generates stars randomly
from a 2-dimensional color-magnitude grid specified by the user.
We chose a grid such that $17 \le {\rm F606W} \le 28$, 
$0 \le {\rm F606W-F814W} \le 3$.
These were placed and solved for one at a time on each set of images
so that no additional crowding is introduced.
The XY position of each artificial star is chosen randomly, but
weighted towards regions with the highest stellar densities in order
to best represent the real measurement conditions. 
A subset of these artificial stars from each
frame (between 15,000 and 20,000 per frame) was chosen for comparison 
with the real stars based on the criterion that their input F606W-F814W
color was within 0.1 mag of the main-sequence fiducial line
(see section 4).   

Charge transfer efficiency corrections were made as described in
\citet{Dolphin2000b}. Aperture corrections to the PSF photometry were
made using 150 - 200 bright and relatively isolated stars on each chip
of each image.  The aperture corrections were typically less than 0.01
mag but were as high as $\sim$ 0.05 mag for the chips sampling the
core of the cluster.

\begin{figure}
\epsscale{0.7}
\plotone{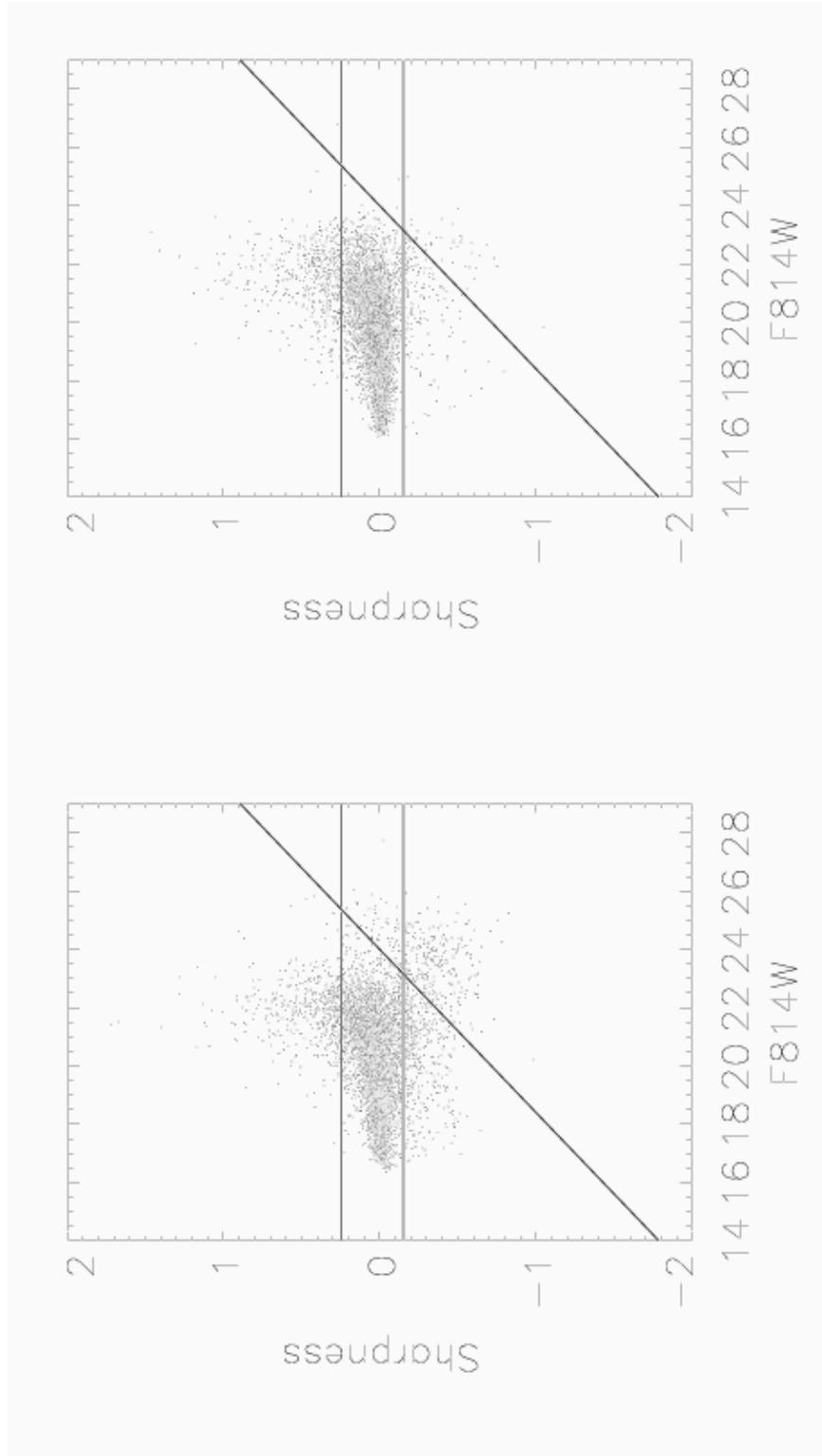}
\caption{\label{fig:sharpness}
Sharpness criteria for stars in the WF3 chip of pointing 3
between 60'' and 120'' from the cluster center. The left hand panel
shows the observed data and the right hand panel
shows the simulated artificial data.}
\end{figure}

The selection of the final star lists for further analysis was made by
imposing a minimum signal-to-noise threshold of 10.0 and making further
cuts using sharpness criteria on a chip-by-chip basis.  The sharpness
reported by HSTPHOT is defined in \citet{Dolphin2000a}.  A perfectly-fit
star has a sharpness of zero, with positive sharpness for stars with a
sharper PSF than this, and negative for objects with a broader
profile. A completely flat profile has a sharpness value of -1.  A
typical example of selection by object sharpness is shown in
Fig.~\ref{fig:sharpness} for the WF3 chip of pointing 3. The sharpness
of all the detected objects found between 60'' and 120'' from the
cluster center with S/N $> 10$ is plotted against F814W magnitude.  (We
will use this same sample field for illustrative purposes throughout the
paper.)  The left-hand panel shows the real data, the right-hand panel
the artificial stars. Selection criteria are made with reference to the
measured sharpness of the artificial stars. The horizontal cuts are made
to reject those stars with poorly-fitting PSFs, the inclined cut is
chosen to reject objects found with low sharpness at fainter magnitudes
that do not appear in the artificial star set.  Some of these faint
detections excluded because of their high negative sharpness are image
artifacts, mainly lying on diffraction spikes from saturated
stars. Others are believed to be blends of faint stars.  The adopted
sharpness cuts for all field/CCD combinations are given in 
Table~\ref{table:CCDfields}.

\begin{figure}
\epsscale{0.7}
\plotone{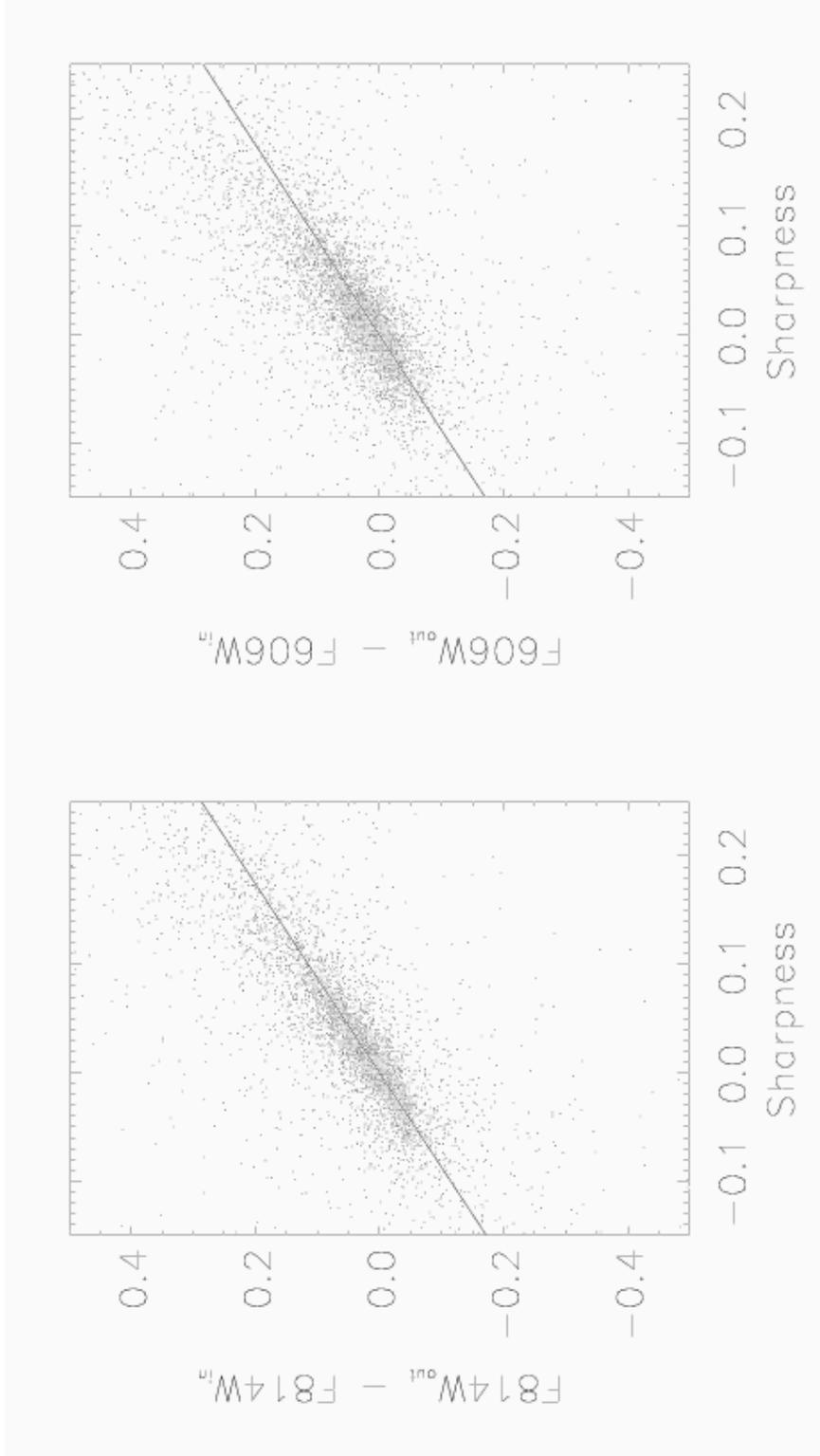}
\caption{\label{fig:sharpmag}
Difference between input and output magnitudes as a 
function of sharpness for the artificial stars in the sample field.}
\end{figure}

A further correction to the derived F814W and F606W WFPC2 flight
system magnitudes was made to correct a trend with sharpness noticed
in the artificial star data. Fig.~\ref{fig:sharpmag} shows the
difference between input and output magnitudes plotted against
sharpness for the artificial stars from the same pointing-3 WF3 field
as above. This effect is present (with the same slope) for all fields,
but as we look
farther away from the core the proportion of stars with non-zero
sharpness decreases and thus it becomes much less significant.
The proportion of stars with non-zero sharpness is also
much greater for fainter stars.
The origin of the effect can be understood as being due
to extreme crowding in the central regions of the cluster. In effect,
the background is not the true sky but rather a lumpy morass of 
undetected stars. The center of a faint, undetected star is more
likely to lie in the wings of a brighter (detected) star then on its
central pixel, leading it to be measured as being brighter and
less sharp. Conversely, a local minimum in the background under a detected star
will most likely result in it being measured as being sharper but with
a smaller flux. 
To verify this, we have performed tests in which we have 
replaced all pixel values below a certain threshold with a constant 
background value, thus reducing the lumpiness of the background. 
Artificial stars were then added to the frame
in the usual way. 
The proportion of the artificial stars subject
to the effect was found to decrease markedly as this threshold was 
increased. 
Since the effect will have influenced all our measurements, the 
real-star magnitudes were corrected to zero sharpness
based on the indicated linear fits to the artificial star data.

\section{Luminosity Function}

\begin{figure}
\plotone{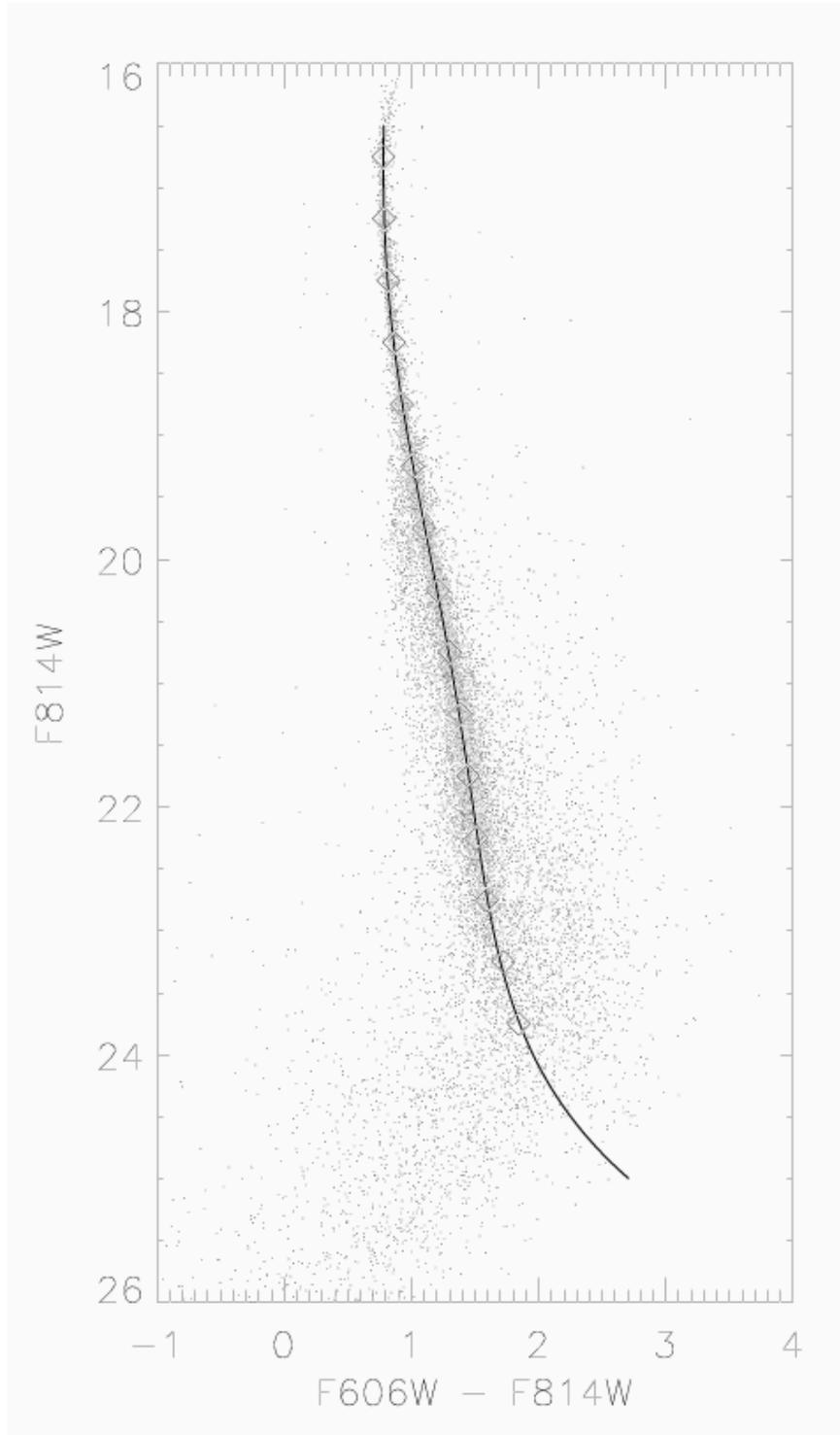}
\caption{\label{fig:CMcombined}
Combined color-magnitude diagram from the 4 PC chips with the main
sequence fiducial from a fifth-order polynomial fit.}
\end{figure}

The combined color-magnitude diagram from the 4 PC chips (one from each
pointing) is shown
in Fig.~\ref{fig:CMcombined}. All stars with $S/N > 4$ and $|sharpness|
< 0.1$ are included. The S/N threshold was deliberately set to
be lower here than what would ultimately be used for our star
counts because we wanted to ensure that our main sequence fiducial
extended to a fainter limiting magnitude. 
The adopted main-sequence fiducial is a
fifth-order polynomial fit to the median F606W-F814W color in 
each 0.5-mag F814W band in the range $16.5 < {\rm F814W} < 24$. 
A 2.5-$\sigma$ clipping routine was used to reject points with outlying
colors in each F814W band before each median color was computed.

\begin{figure}
\epsscale{0.7}
\plotone{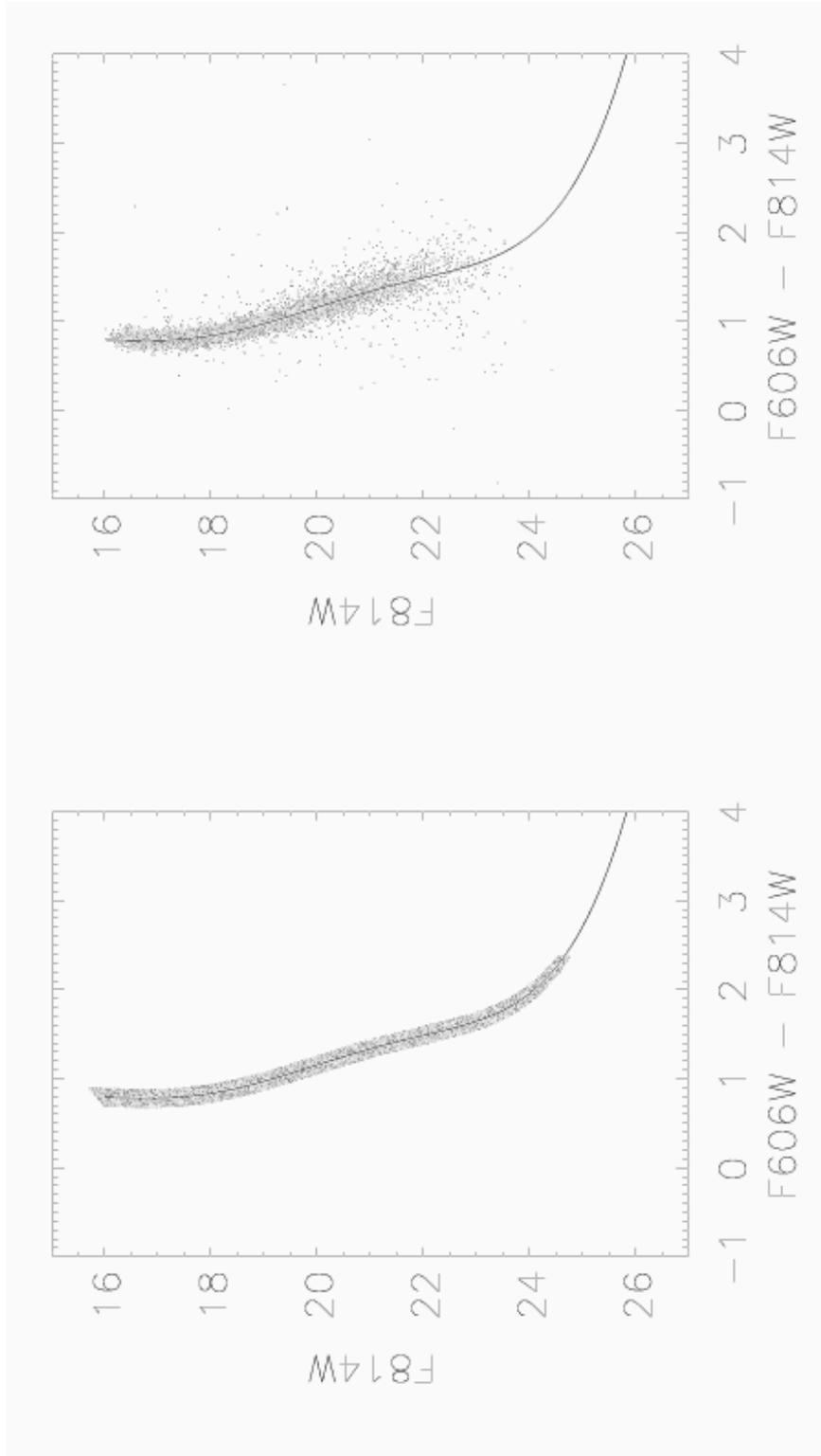}
\caption{\label{fig:CMartificial}
Input (left) and output (right) artificial star color-magnitude diagrams for WF3
pointing-3 field and between 60'' and 120'' from the cluster center.}
\end{figure}

In order that our artificial star tests might best represent the
actual colors and magnitudes of the measured stars, we selected
only those artificial stars whose input magnitudes fell within
0.1 mag in color from the calculated main sequence fiducial.  
Sample input and output color-magnitude diagrams for the artificial
stars in our sample field are shown in Fig.~\ref{fig:CMartificial}.

Since we are interested in determining how the luminosity function of
M22 varies as a function of radius from the cluster center, the sets
of real and artificial stars for each CCD were divided into concentric
annular bins. These annuli were initially chosen at 60'' radial
intervals extending from the center of the cluster out to 300'' as
shown in Fig.~\ref{fig:m22area}. These 16 CCD fields and 5 radial bins
thus give a grid of 80 possible luminosity functions to be
calculated. In practice, at most two of these radial bins are well 
sampled by a given CCD. In order to better sample the core, 
we repeated our analysis using 20'' annuli of which only the innermost
five contained sufficient numbers of stars for 
luminosity functions to be computed with any degree of significance.

\begin{figure}
\epsscale{0.7}
\plotone{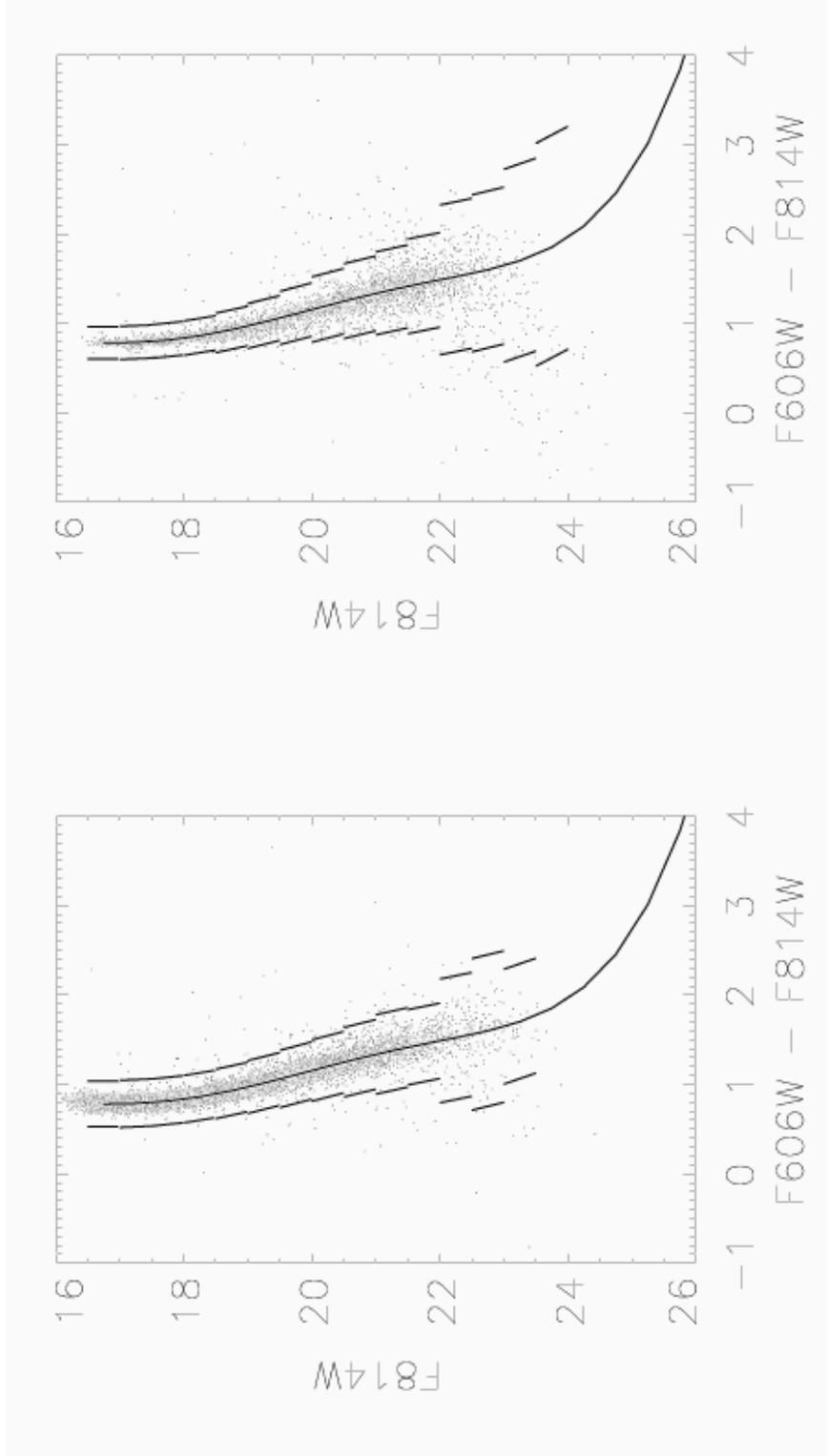}
\caption{\label{fig:CMclip}
Artificial (left) and real star (right) color-magnitude diagrams for the sample field
showing the main sequence fiducial and clipping curves
used for statistical adjustment of star counts for field-star contamination.}
\end{figure}

In Fig.~\ref{fig:CMclip} we show the color-magnitude diagram for the
sample pointing-3, WF3, 60-120'' bin. The left panel shows the real star
photometry, the right panel is for the artificial stars. Indicated is
the main-sequence fiducial (calculated as described above from the 
real-star data for all PC
fields) and two 2.5-$\sigma$ curves used for statistically correcting the star
counts for field-star contamination. 
Unfortunately the field-star
densities of \citet{Ratnatunga1985} do not extend to galactic
latitudes as near to the Plane as M22 ($b = -7.55$).  The selection
curves were calculated from the artificial stars as follows.  First,
the fiducial main sequence color was subtracted from each point.  The
resultant $\Delta$(F814W-F606W) values were then subjected to an
iterative 2.5-$\sigma$ clipping algorithm,
for each 0.5-mag F814W bin and the fiducial main-sequence color added
back to the two 2.5-$\sigma$ limits. 
Thus, in
the absence of field-star contamination, 98.75\% of main-sequence stars
are found between the selection curves. Equivalently, the number of
stars outside the selection lines should be 1.26\% of the number
inside.  To estimate field star contamination, we count the number of stars
inside and outside the selection lines in each 0.5-mag F814W bin
within the color range $-1 < {\rm F814W-F606W} < 4$. If the outside
count is greater than 1.26\% of the inner count then we adjust the
inner count downwards by the excess, weighted for the differing
color-ranges covered. Exactly the same algorithm is applied to the 
artificial-star data and to the real stars.

The application of the 2.5~$\sigma$ clipping criterion provides
us with an upper limit to the luminosity function in that magnitude
bins along the main sequence, although clipped to 5~$\sigma$ in color,
will also contain background Galactic bulge stars. The bulge
color-magnitude diagram \citep{Holtzman1998} overlaps that of M22
and its luminosity function increases with magnitude.

The luminosity function of the cluster $\phi$ is
defined by
\begin{equation}
dN(M) = \phi(M) dM,
\end{equation}
where $dN(M)$ is the number of stars per unit area with magnitudes
between $M$ and $M + dM$.
In each 0.5-mag F814W bin,
$\phi_{i}$, is related to the measured star counts, $n_{i}$, by the
equation
\begin{equation}
T.{\underline \phi} = {\underline n}
\end{equation}
\citep{Drukier1988}.  The element $T_{ij}$ of the photometric
completion matrix,
$T$, represents the probability that a star from magnitude bin $j$
will be measured in magnitude bin $i$.  This matrix is constructed
from the artificial star counts by comparing each measured F814W
magnitude with its input magnitude.  For the case of perfect photometry
with no ``bin jumping'', the matrix is diagonal. In practice, there is a
small probability, increasing towards fainter magnitudes, that a given
star is scattered up or down in luminosity.

We decided to only measure luminosity functions where the diagonal
matrix element was greater than 30\%. Experiments showed that
constructing the matrix with a limiting magnitude 2 bins below this
level was sufficient to assess contamination from fainter stars that
have scattered upwards, but not so faint as to cause the matrix to be
ill conditioned. The mean photometric completeness in the lowest bin
for all our field/annulus combinations was 0.45.  One bin above the
cutoff, the mean photometric completess was 0.56.  In calculating the
luminosity function we took into account Poisson errors in the star
counts for $n$ and also for the artificial star data in the matrix
$T$.

\begin{figure}
\epsscale{1.0}
\plottwo{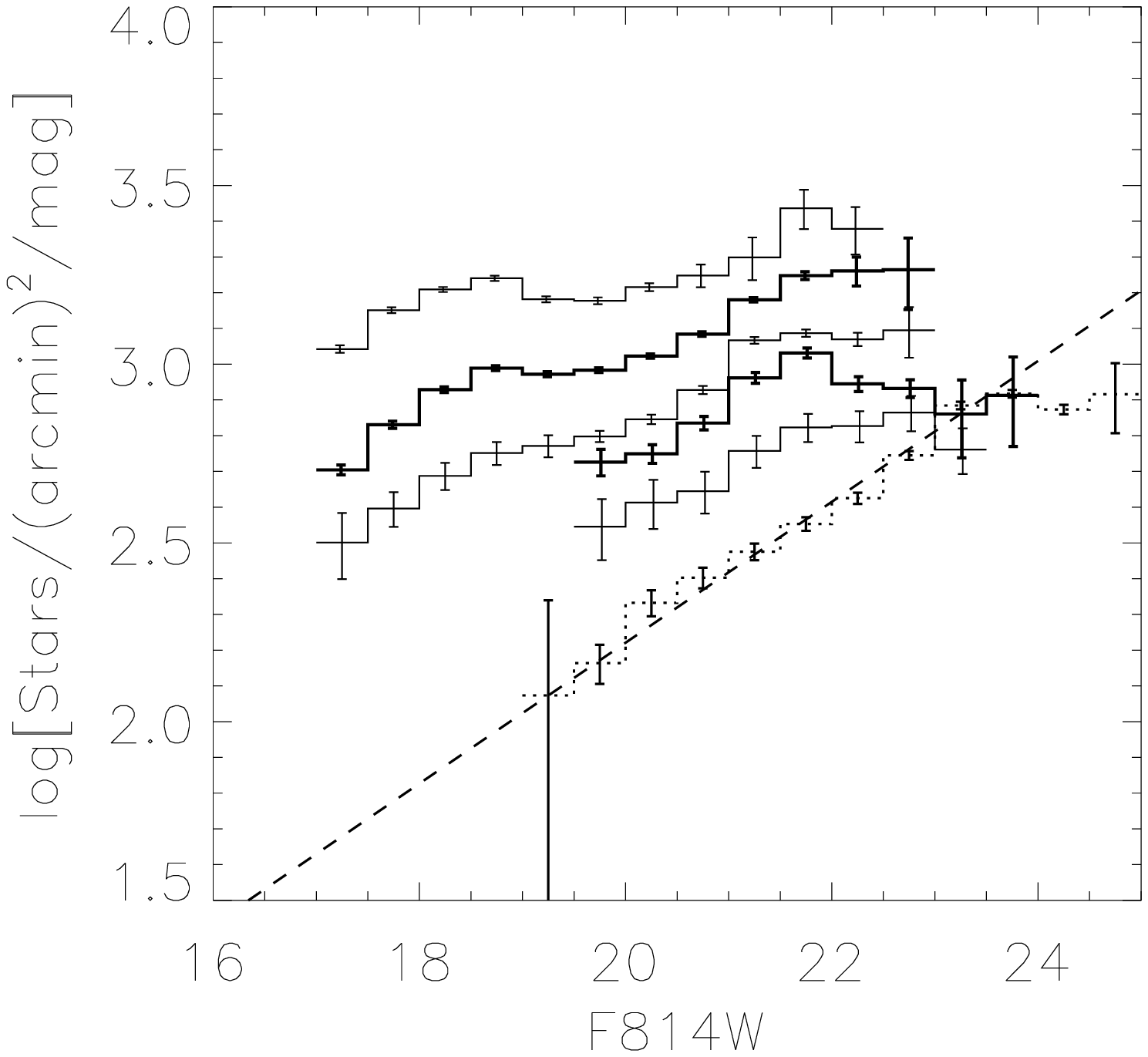}{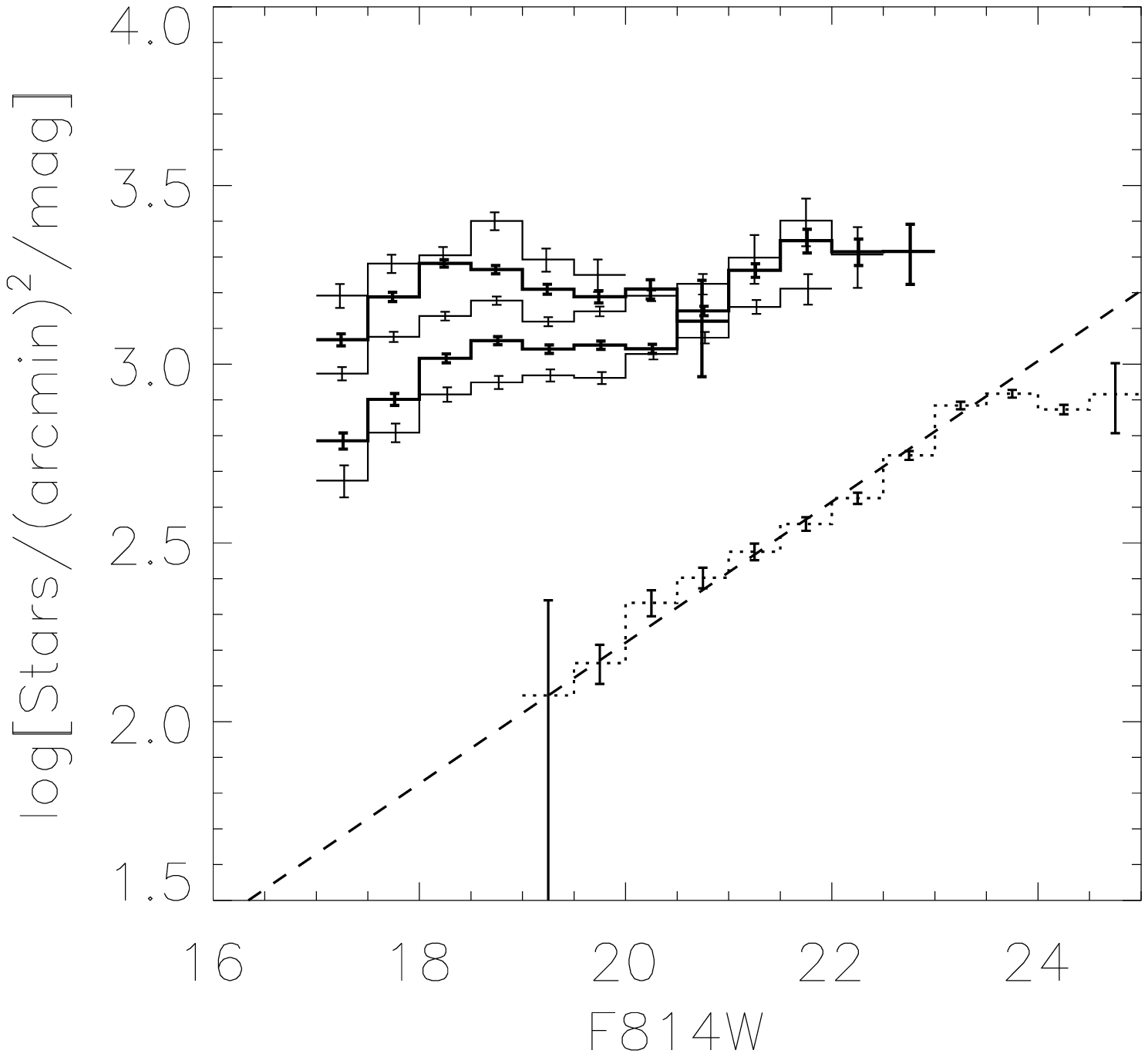}
\caption{\label{fig:Lumf}
Luminosity function for concentric annular bins from the center of the
cluster with the uppermost curve in each panel being for the central
circular bin. The lowest curve in each panel (dotted line) is
the bulge luminosity function measured from an archival exposure
offset from M22. A linear fit to the bulge luminosity function
is indicated with a dashed line.
The left panel is for annular bins of 60'' radial increment
and the right hand panel is for the innermost five annuli with
a 20'' radial increment.}
\end{figure}

A final scale correction to the derived luminosity functions is made
to allow for the spatial area sampled and the 0.5-mag F814W bin
size. 
The individual luminosity functions for the different
chip/radius combinations were statistically combined
into luminosity functions for each radial bin and
the 
combined luminosity functions from the various fields at different
radii from the center of the cluster are given in
Tables~\ref{table:lumfcomb60}-\ref{table:lumfcomb20} and plotted in
Fig.~\ref{fig:Lumf}.

Also shown in Fig.~\ref{fig:Lumf} is a luminosity function we have
derived for the Galactic bulge local to M22. For this calculation
we used the WFPC2 archival dataset u27xjd01t. This is a single,
non-CRSPLIT,  2400 s F814W exposure of a field offset from the center
of M22 by approximately 9 arcmin. The four WFPC2 CCD frames from 
this exposure were processed in the same way as the M22 observations.
Artificial star tests were again used to correct the derived
luminosity functions for photometric completeness and the corrected
luminosity functions for the four chips were statistically combined. 
The photometric completeness
for all chips was around 75\% at F814W = 22 and 50\% at F814W = 24.

\begin{figure}
\epsscale{1.0}
\plottwo{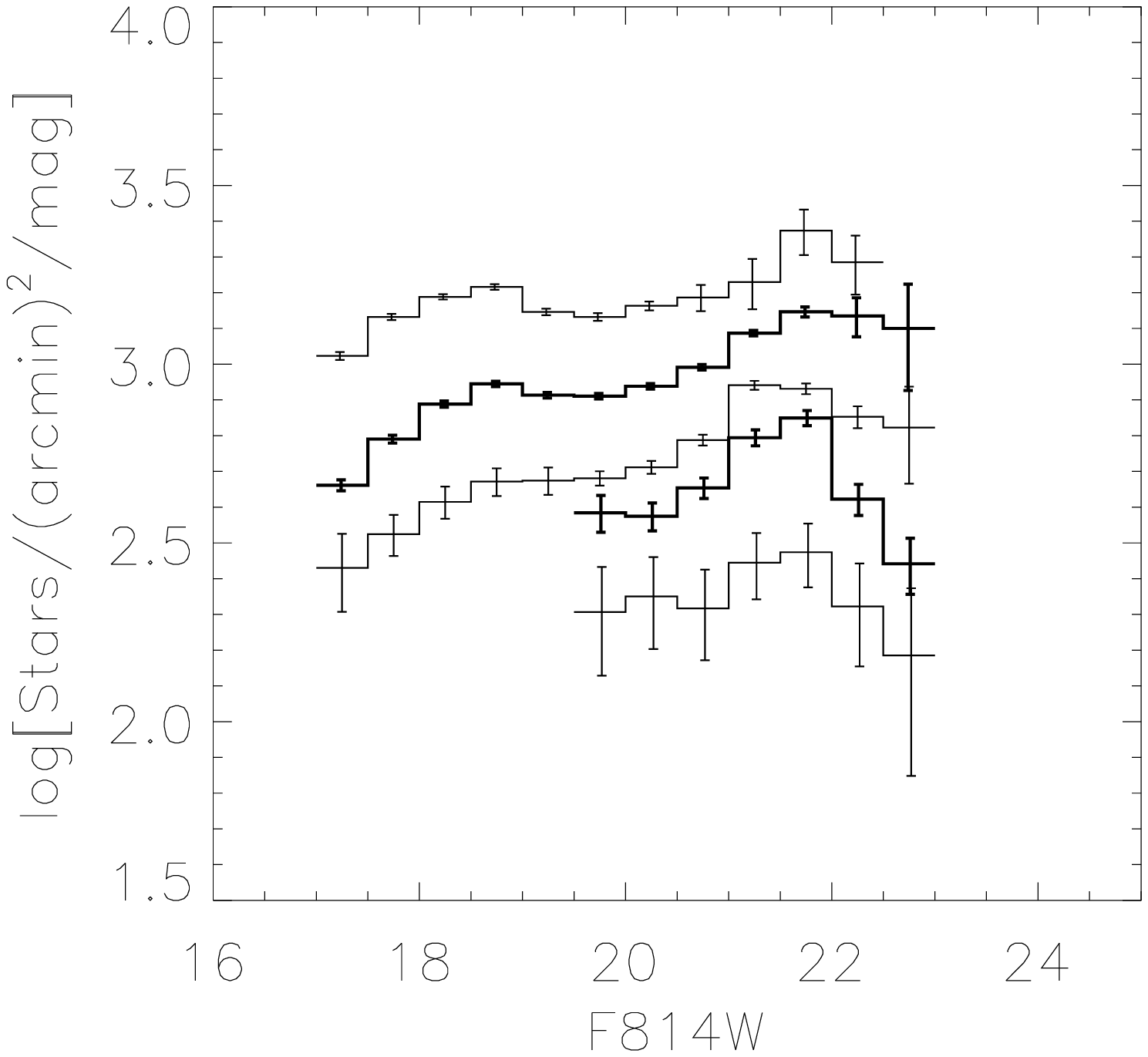}{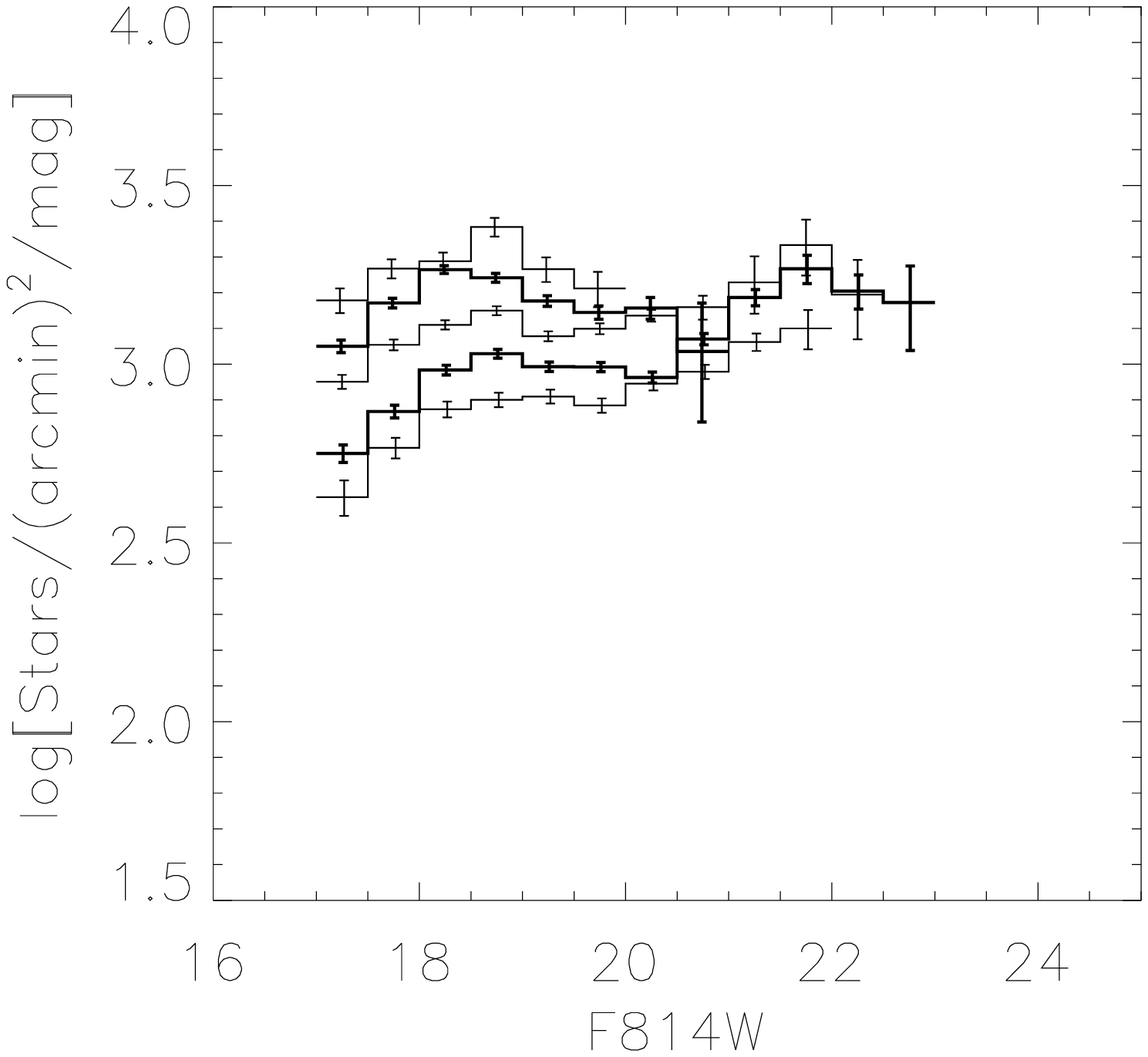}
\caption{\label{fig:Lumfcorr}
Luminosity functions from Fig.~\ref{fig:Lumf} that have had the 
background Galactic bulge luminosity function subtracted.}
\end{figure}

Since the derived bulge luminosity function is approximately
linear over $19 < {\rm F814W} < 23$ we have made a weighted linear
fit to the bulge luminosity function in this region,
$\log N = 0.197 {\rm F814W} - 1.72$.  
Comparison with Fig. 5 of \citet{Holtzman1998} shows that the 
Baade's window luminosity function is also linear in
this region (assuming the same distance and extinction)
and has a similar slope.
We have corrected our M22 luminosity functions for background
bulge contamination by subtracting the indicated linear
fit extrapolated to brighter magnitudes. Again referring to
Fig. 5 of \citet{Holtzman1998}, the Baade's window luminosity
function drops more rapidly for magnitudes brighter than $M_{I} = 3.25$ 
(F814W = 18.5) suggesting we may have over-corrected the brighter
magnitudes. However, this over-correction is at most 0.05 in
the log luminosity function. Our resulting corrected luminosity
functions for M22 are given in
Tables~\ref{table:lumfcomb60corr}-\ref{table:lumfcomb20corr}
and shown in Fig.~\ref{fig:Lumfcorr}.

\section{Mass function}

\begin{figure}
\epsscale{1.0}
\plotone{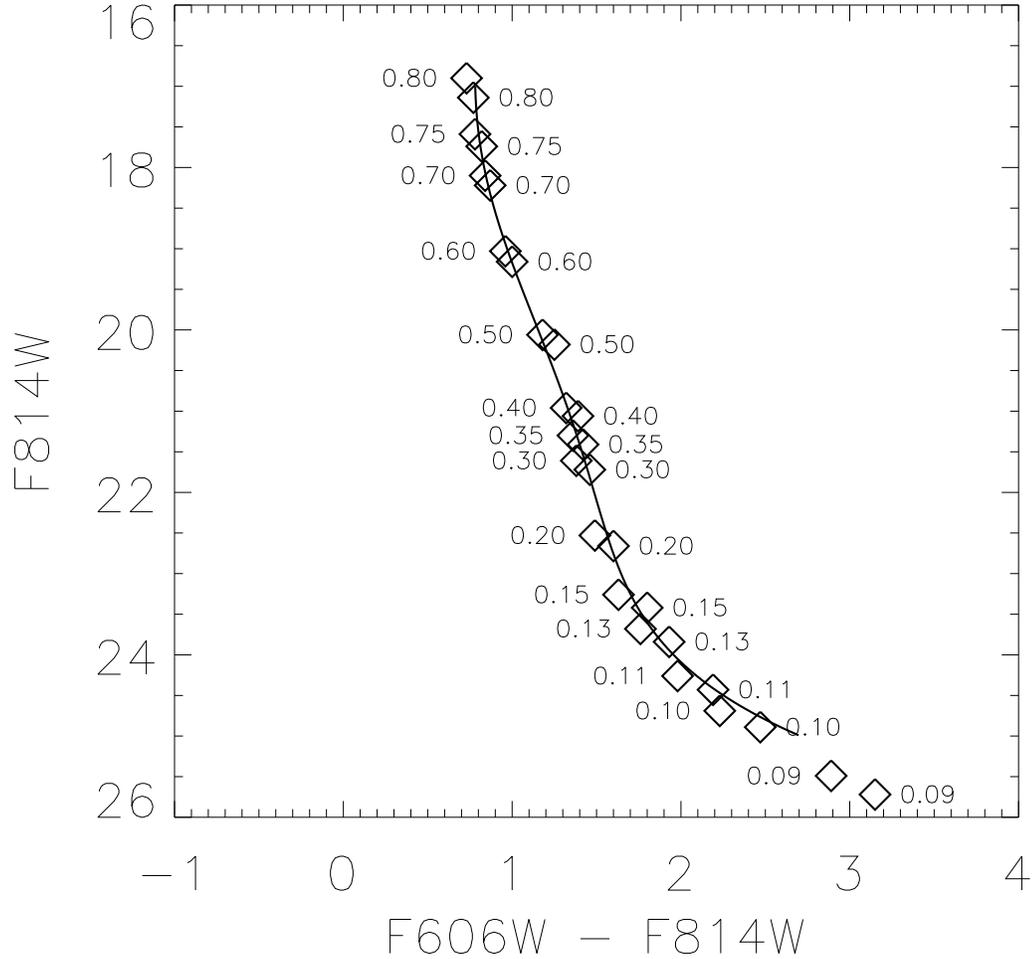}
\caption{\label{fig:CMDiso}
Observed main-sequence fiducial (solid line) with models
of \citet{Baraffe1997} for [M/H] = -1.3 (numbered to left) and 
[M/H] = -1.0 (numbered to right) 
with masses indicated. The model points have been transformed to the
observational plane assuming $(m-M)_{V} = 13.6$ and $E(B-V) = 0.34$
from \citet{Harris1996} and the extinction coefficients of
\citet{Schlegel1998}.}
\end{figure}

To transform the luminosity functions into mass functions we
use the 10-Gyr evolutionary models of \citet{Baraffe1997} for 
metal-poor low-mass stars. These models have been shown to 
be a good fit to 
the lower main sequences of globular clusters observed
by HST and the
authors have made available tables of mass vs luminosity
in the WFPC2 flight system filter set.
We follow \citet{Baraffe1997} and calculate [M/H] following 
the prescription of \citet{Ryan1991} for halo subdwarfs.
For the metallicity range of interest, [M/H] $\approx$ [Fe/H] + 0.35.
\citet{Harris1996} lists [Fe/H] = $-1.64$ for M22 while \citet{Caretta1997}
found [Fe/H] = $-1.48$ $\pm$ 0.03.
In Fig.~\ref{fig:CMDiso} we thus compare the
main-sequence fiducial of M22 with that predicted by
the models for [M/H] = $-1.3$ and [M/H] = $-1.0$.
We have transformed the model points to the observational plane using
$(m-M)_{V} = 13.60$ and $E(B-V) = 0.34$ (again from \citet{Harris1996})
and taken the relative extinction coefficients for the WFPC2 filters from
\citet{Schlegel1998}. The colors and luminosities for both models provide
a remarkable match to our photometric main sequence fiducial. We adopt the
relation for [M/H] = $-1.0$ as the match is slightly better to both the 
photometry and the (presumably more accurate) Caretta \& Gratton metallicity.

\begin{figure}
\epsscale{1.5}
\plottwo{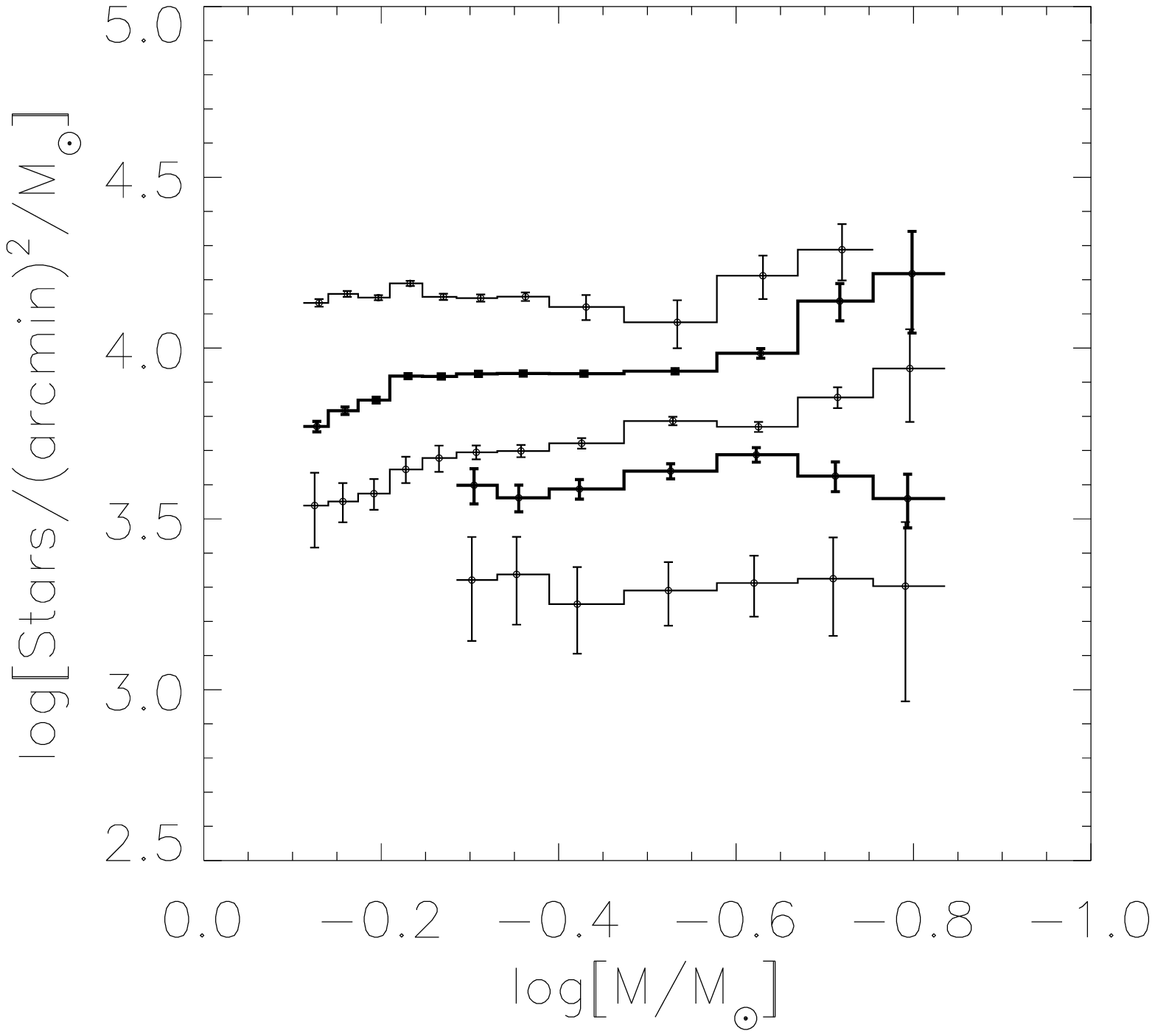}{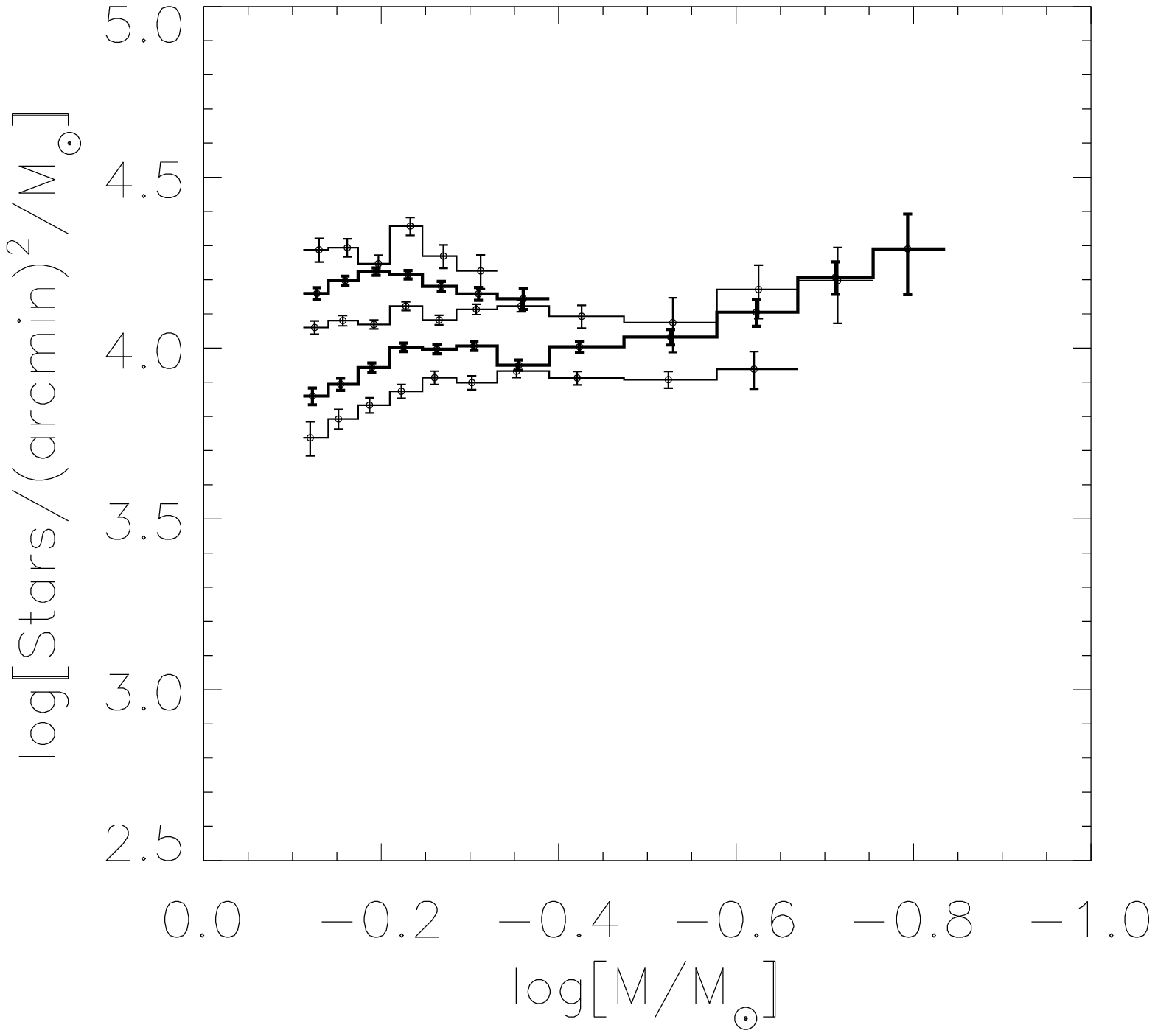}
\caption{\label{fig:massf}
Mass function for concentric annular bins from the center of the
cluster with the uppermost curve in each panel being for the central
circular bin. The left panel is for annular bins of 60'' radial increment
and the right hand panel is for the innermost five annulli with
a 20'' radial increment.}
\end{figure}

The mass function $\zeta(m)$, defined by
\begin{equation}
dN(m) = \zeta(m) dm
\end{equation}
where $dN(m)$ is the number of stars per unit area with
masses between $m$ and $m + dm$,
 is related to the luminosity 
function $\phi({\rm F814W})$ by
\begin{equation}
\zeta(m) dm = \phi({\rm F814W}) d{\rm F814W}.
\end{equation}
The mass-luminosity relation from the
theoretical [M/H] = -1.0 isochrone was thus used to assign
a mass range to each F814W bin. The derivative 
of the relation at the center of each bin
was used to
translate  the luminosity functions to the mass functions
shown in Fig.~\ref{fig:massf} and listed in Tables~\ref{table:mass60}-\ref{table:mass20}.

The mass functions for the annular bins can be characterised
by examining three regions, $\log m \lesssim -0.6$, 
$-0.6 \lesssim \log m \lesssim -0.3$ and
$\log m \gtrsim -0.3$. For $\log m \lesssim -0.6$, the mass functions
interior to a 180'' radius
rise towards lower masses with an approximately constant power
law index $\alpha \approx 
1.0$ to 1.3, where $\zeta(m) \propto m^{-\alpha}$.
Our data do not extend to faint enough
magnitudes to see any turnover in these mass functions.
Between $\log m \approx -0.6$ and $\log m \approx -0.3$
the mass functions are flat ($\alpha \approx 0$).
Clear evidence of mass segregation is seen
for $\log m \approx -0.3$. Outside of approximately $r_{c}$
(60'' - 85''), the mass function decreases with increasing mass
($\alpha \approx 1.2$).
Within the core and towards the center, there is an increasing tendancy
for the mass function to flatten and then rise towards higher masses, as 
illustrated in the mass functions for 20'' annular bins.

\section{Simulation of Dynamical Structure}

Having derived the spatially resolved mass function for NGC 6656, we
next address the issue as to whether the degree of mass segregation can
be accounted for by the theory of relaxation.  To study the dynamical
properties of the cluster, we have employed the multi-mass Michie--King
models originally developed by \citet{Meylan1987, Meylan1988} and later
suitably modified by \citet{Pulone1999} and \citet{DeMarchi2000} for the
general case of clusters with a set of radially varying luminosity
functions. 
Each model run is characterised by a mass function (MF) in the form of
an exponential $dN/d \log m
\propto m^{-x}$, with a variable exponent $x$ (note that $\alpha = 1 +
x$), and by four structural
parameters describing, respectively, the scale radius ($ r_{\rm c}$),
the scale velocity ($v_{\rm s}$), the central value of the dimensionless
gravitational potential ($W_{\rm o}$) and the anisotropy radius ($r_{\rm
a}$). From the parameter space defined in this way, we have selected
those models that simultaneously fit both the observed surface
brightness (SBP) and velocity dispersion (VDP) profiles of the cluster
as measured, respectively, by \citet{Trager1995} and
\citet{Peterson1994}. The fit to the SBP and VDP, however, can only
constrain $r_{\rm c}$, $v_{\rm s}$, $W_{\rm o}$, and $r_{\rm a}$ while
still allowing the MF to take on a variety of shapes. To break this
degeneracy, we further impose the condition that the model MF agree with
the observed LF.

Since Michie--King modeling only provides a ``snapshot'' of the current
dynamical state of the cluster, one finds it useful to define the
global mass function (GMF), the mass distribution of all cluster
stars at present, as the MF that the cluster would have simply as a
result of stellar evolution (that is, ignoring any local modifications
induced by internal dynamics and/or the interaction with the Galactic
tidal field). Clearly, in this case the IMF and GMF of main sequence 
(un-evolved)
stars is the same. For practical purposes, the GMF has been divided
into sixteen different mass classes, covering main sequence stars, 
white dwarfs,
and heavy remnants, precisely as described in \citet{Pulone1999}.

Our parametric modelling approach assumes energy equipartition amongst
stars of different masses. Thus, we have run a large number of trials
to see whether we could find a set of parameters for the GMF (i.e. a
suitable GMF ``shape'') such that the local MFs produced by mass
segregation would locally fit the observations. Our exercise confirms
what we had already implicitly shown in Fig.~\ref{fig:massf} and described
above:  as long as a single value of the exponent $x$ is used for the
GMF over the mass range $0.2 - 0.8$\,M$_\odot$, none of the predicted MF
can be fitted to our data. In fact, a change of slope is needed at $m
\simeq 0.4$\,M$_\odot$ so that both the flat and rising portions of the
local MF can be reproduced. If we then allow the MF to take on more
than one slope, the GMF that best fits the observations is one with
$x=0.2$ ($\alpha = 1.2$) for stars in the range $0.4 - 0.8$\,M$_\odot$
and $x=-0.5$ ($\alpha = 0.5$) at
smaller masses. 

Although stars more massive than $\sim 0.8$\,M$_\odot$ have evolved and
are no longer visible, the shape of the IMF in this mass range has
strong implications as to the fraction of heavy remnants in the cluster
and, as such, on the central velocity dispersion. We find that a value
of $x=0.9$ ($\alpha = 1.9$) for stars in the range $100 - 0.8$\,M$_\odot$ gives the best
fit to the data and to the cluster's structural parameters as given in
the literature. The latter, along with those of our best fitting model,
are presented in Table~\ref{table:modelparams}. The agreement is
excellent, apart from a small difference in the value of the core
radius. We note here that global cluster MF is shallower than Salpeter's
IMF, which would have $x=1.35$. The total implied cluster mass is $2.7
\times 10^5$\,M$_\odot$ and the mass-to-light ratio is on average $m/L =
1.6$, with $m/L \simeq 2$ in the core. These are all very typical values
for a cluster of this type and confirm that the observed degree of mass
segregation is indeed what would be expected from dynamical relaxation.

\section{Summary}

Extensive HST imaging of M22 has been used to determine
the luminosity function for this globular cluster at a
number of different radii from the cluster center. 
Using the \citet{Baraffe1997} stellar isochrones, we have transformed
these luminosity functions into mass functions.
The proportion of higher-mass stars was found to be significantly
enhanced within one core radius of the center of the cluster
compared to regions outside the core.
This
is the first time that such a detailed mapping of mass
segregation from the mid main sequence to the turnoff 
has been performed for a globular cluster.
 
Numerical simulation of the radial mass spectrum of M22
using multi-mass King-Michie models has shown that the degree
of mass segregation found is well predicted by the standard theory
of cluster relaxation.

 
\clearpage
 
\clearpage
 
\clearpage

\clearpage
 
\clearpage
 
\clearpage

\clearpage
 
\clearpage

\clearpage

\clearpage

\clearpage
\begin{figure}
\epsscale{1.0}
\plotone{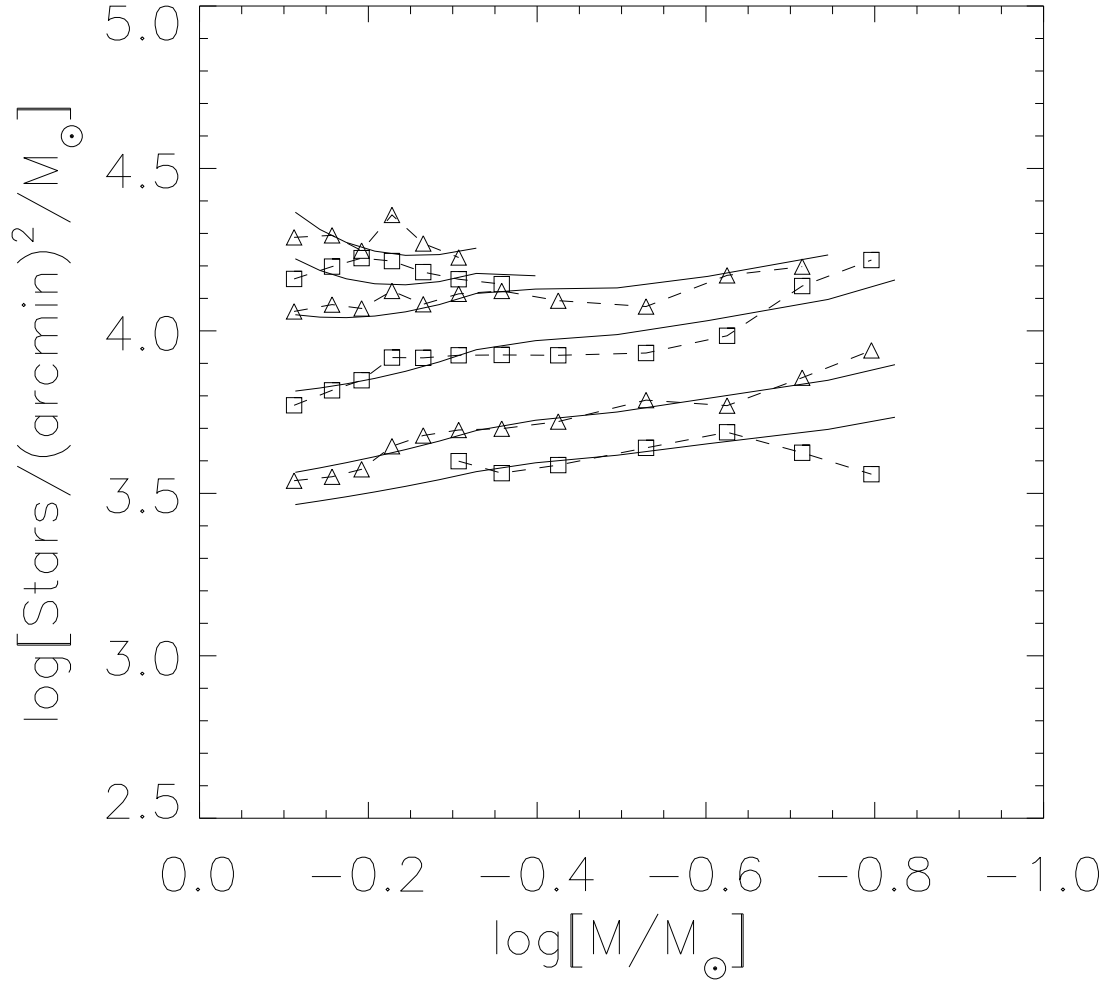}
\caption{\label{fig:model}
Observed MFs (boxes) and the MFs predicted by the model (solid lines) at
the radial distances 14", 32", 51", 95", 153"
and 212". These distances are at the geometric mean of each measured 
annulus. Error bars are not plotted as they are comparable
or (usually) smaller than the size of the symbols.}
\end{figure}


\clearpage
\begin{deluxetable}{|c|c|c|c|c|c|c|}
\tabletypesize{\small}
\tablecaption{\label{table:CCDfields} CCD fields and sharpness cut
criteria}
\tablehead{
\colhead{Field Name}  & \colhead{Pointing} & \colhead{CCD} & 
\multicolumn{4}{|c|}{Sharpness cut criteria} \\
 & & & \colhead{Min} & \colhead{Max} & \colhead{Slope} & \colhead{Zero point}
}
\startdata
1   &  1  & PC1 & -0.15 & 0.20 & 2.5 & -4.29 \\
2   &  1  & WF2 & -0.15 & 0.25 & 2.5 & -4.29 \\
3   &  1  & WF3 & -0.15 & 0.25 & 2.5 & -4.29 \\
4   &  1  & WF4 & -0.15 & 0.25 & 2.5 & -4.29 \\
5   &  2  & PC1 & -0.15 & 0.20 & 2.5 & -4.29 \\
6   &  2  & WF2 & -0.15 & 0.25 & 2.5 & -4.29 \\
7   &  2  & WF3 & -0.15 & 0.25 & 2.5 & -4.29 \\
8   &  2  & WF4 & -0.15 & 0.25 & 2.5 & -4.29 \\
9   &  3  & PC1 & -0.15 & 0.20 & 2.5 & -4.29 \\
10  &  3  & WF2 & -0.15 & 0.25 & 2.5 & -4.29 \\
11  &  3  & WF3 & -0.15 & 0.25 & 2.5 & -4.29 \\
12  &  3  & WF4 & -0.15 & 0.25 & 2.5 & -4.29 \\
13  &  4  & PC1 & -0.15 & 0.20 & 2.5 & -4.46 \\
14  &  4  & WF2 & -0.15 & 0.25 & 2.5 & -4.46 \\
15  &  4  & WF3 & -0.15 & 0.25 & 2.5 & -4.46 \\
16  &  4  & WF4 & -0.15 & 0.25 & 2.5 & -4.46 
\enddata
\end{deluxetable}

\clearpage
\begin{deluxetable}{c|rr|rr|rr|rr|rr|}
\rotate
\tabletypesize{\footnotesize}
\tablecaption{\label{table:lumfcomb60} Combined Luminosity functions for
  60'' annular bins}
\tablehead{
\colhead{Radius}  & \multicolumn{2}{|c|}{0--60} & \multicolumn{2}{|c|}{60--120} 
& \multicolumn{2}{|c|}{120--180} & \multicolumn{2}{|c|}{180--240} & \multicolumn{2}{|c|}{240--300} \\
\colhead{F814W}  & \colhead{$\phi$} & \colhead{$\sigma_{\phi}$}  
 & \colhead{$\phi$} & \colhead{$\sigma_{\phi}$} & \colhead{$\phi$} & \colhead{$\sigma_{\phi}$}
 & \colhead{$\phi$} & \colhead{$\sigma_{\phi}$} & \colhead{$\phi$} & \colhead{$\sigma_{\phi}$}
}
\startdata
17.0-17.5 &  1103 &    27 &   506 &    16 &   317 &    66 &       &       &       &       \\
17.5-18.0 &  1416 &    27 &   677 &    16 &   395 &    44 &       &       &       &       \\
18.0-18.5 &  1618 &    26 &   849 &    15 &   487 &    43 &       &       &       &       \\
18.5-19.0 &  1739 &    29 &   975 &    15 &   564 &    42 &       &       &       &       \\
19.0-19.5 &  1519 &    29 &   937 &    15 &   591 &    42 &       &       &       &       \\
19.5-20.0 &  1504 &    33 &   963 &    14 &   628 &    22 &   532 &    45 &   351 &    68 \\
20.0-20.5 &  1643 &    42 &  1054 &    15 &   701 &    21 &   562 &    34 &   410 &    64 \\
20.5-21.0 &  1770 &   129 &  1214 &    17 &   847 &    22 &   684 &    30 &   441 &    59 \\
21.0-21.5 &  1991 &   272 &  1514 &    21 &  1166 &    25 &   916 &    32 &   572 &    59 \\
21.5-22.0 &  2731 &   344 &  1769 &    45 &  1222 &    29 &  1075 &    35 &   666 &    60 \\
22.0-22.5 &  2390 &   362 &  1825 &   170 &  1174 &    50 &   881 &    42 &   671 &    67 \\
22.5-23.0 &       &       &  1838 &   415 &  1244 &   201 &   856 &    49 &   732 &    83 \\
23.0-23.5 &       &       &       &       &       &       &   725 &   178 &   577 &    84 \\
23.5-24.0 &       &       &       &       &       &       &   818 &   230 &       &         
\enddata
\end{deluxetable}

\clearpage
\begin{deluxetable}{c|rr|rr|rr|rr|rr|}
\rotate
\tabletypesize{\footnotesize}
\tablecaption{\label{table:lumfcomb20} Combined Luminosity functions for
  20'' annular bins}
\tablehead{
\colhead{Radius}  & \multicolumn{2}{|c|}{0--20} & \multicolumn{2}{|c|}{20--40} 
& \multicolumn{2}{|c|}{40--60} & \multicolumn{2}{|c|}{60--80} & \multicolumn{2}{|c|}{80--100} \\
\colhead{F814W}  & \colhead{$\phi$} & \colhead{$\sigma_{\phi}$}  
 & \colhead{$\phi$} & \colhead{$\sigma_{\phi}$} & \colhead{$\phi$} & \colhead{$\sigma_{\phi}$}
 & \colhead{$\phi$} & \colhead{$\sigma_{\phi}$} & \colhead{$\phi$} & \colhead{$\sigma_{\phi}$}
}
\startdata
17.0-17.5 &  1556 &   120 &  1171 &    45 &   941 &    40 &   610 &    32 &   473 &    49 \\
17.5-18.0 &  1911 &   113 &  1543 &    46 &  1193 &    40 &   797 &    30 &   644 &    39 \\
18.0-18.5 &  2016 &   113 &  1915 &    46 &  1363 &    39 &  1038 &    30 &   823 &    38 \\
18.5-19.0 &  2515 &   145 &  1840 &    50 &  1506 &    40 &  1164 &    30 &   890 &    37 \\
19.0-19.5 &  1961 &   145 &  1621 &    52 &  1316 &    38 &  1102 &    30 &   931 &    36 \\
19.5-20.0 &  1777 &   184 &  1544 &    60 &  1406 &    44 &  1131 &    30 &   915 &    36 \\
20.0-20.5 &       &       &  1623 &   100 &  1554 &    52 &  1104 &    32 &  1068 &    37 \\
20.5-21.0 &       &       &  1319 &   396 &  1677 &   111 &  1410 &    43 &  1186 &    44 \\
21.0-21.5 &       &       &       &       &  1987 &   309 &  1830 &    80 &  1447 &    65 \\
21.5-22.0 &       &       &       &       &  2522 &   384 &  2216 &   169 &  1627 &   159 \\
22.0-22.5 &       &       &       &       &  2027 &   391 &  2063 &   175 &       &       \\
22.5-23.0 &       &       &       &       &       &       &  2067 &   395 &       &       
\enddata
\end{deluxetable}

\clearpage
\begin{deluxetable}{c|rr|rr|rr|rr|rr|}
\rotate
\tabletypesize{\footnotesize}
\tablecaption{\label{table:lumfcomb60corr} Combined Luminosity functions for
  60'' annular bins after subtraction of the background Galactic bulge
  luminosity function.}
\tablehead{
\colhead{Radius}  & \multicolumn{2}{|c|}{0--60} & \multicolumn{2}{|c|}{60--120} 
& \multicolumn{2}{|c|}{120--180} & \multicolumn{2}{|c|}{180--240} & \multicolumn{2}{|c|}{240--300} \\
\colhead{F814W}  & \colhead{$\phi$} & \colhead{$\sigma_{\phi}$}  
 & \colhead{$\phi$} & \colhead{$\sigma_{\phi}$} & \colhead{$\phi$} & \colhead{$\sigma_{\phi}$}
 & \colhead{$\phi$} & \colhead{$\sigma_{\phi}$} & \colhead{$\phi$} & \colhead{$\sigma_{\phi}$}
}
\startdata
17.0-17.5 & 1055 &    27 &   458 &    16 &   269 &    66 &       &       &       &       \\
17.5-18.0 & 1356 &    27 &   617 &    16 &   335 &    44 &       &       &       &       \\
18.0-18.5 & 1543 &    26 &   774 &    15 &   412 &    43 &       &       &       &       \\
18.5-19.0 & 1645 &    29 &   881 &    15 &   470 &    42 &       &       &       &       \\
19.0-19.5 & 1401 &    29 &   819 &    15 &   472 &    42 &       &       &       &       \\
19.5-20.0 & 1356 &    33 &   814 &    14 &   480 &    22 &   384 &    45 &   203 &    68 \\
20.0-20.5 & 1456 &    42 &   868 &    15 &   515 &    21 &   375 &    34 &   224 &    64 \\
20.5-21.0 & 1537 &   129 &   981 &    17 &   613 &    22 &   451 &    30 &   207 &    59 \\
21.0-21.5 & 1697 &   272 &  1221 &    21 &   873 &    25 &   623 &    32 &   278 &    59 \\
21.5-22.0 & 2364 &   344 &  1402 &    45 &   854 &    29 &   707 &    35 &   298 &    60 \\
22.0-22.5 & 1928 &   362 &  1364 &   170 &   712 &    50 &   419 &    42 &   210 &    67 \\
22.5-23.0 &      &       &  1259 &   415 &   665 &   201 &   276 &    49 &   153 &    83 
\enddata
\end{deluxetable}

\clearpage
\begin{deluxetable}{c|rr|rr|rr|rr|rr|}
\rotate
\tabletypesize{\footnotesize}
\tablecaption{\label{table:lumfcomb20corr} Combined Luminosity functions for
  20'' annular bins after subtraction of the background Galactic bulge
  luminosity function.}
\tablehead{
\colhead{Radius}  & \multicolumn{2}{|c|}{0--20} & \multicolumn{2}{|c|}{20--40} 
& \multicolumn{2}{|c|}{40--60} & \multicolumn{2}{|c|}{60--80} & \multicolumn{2}{|c|}{80--100} \\
\colhead{F814W}  & \colhead{$\phi$} & \colhead{$\sigma_{\phi}$}  
 & \colhead{$\phi$} & \colhead{$\sigma_{\phi}$} & \colhead{$\phi$} & \colhead{$\sigma_{\phi}$}
 & \colhead{$\phi$} & \colhead{$\sigma_{\phi}$} & \colhead{$\phi$} & \colhead{$\sigma_{\phi}$}
}
\startdata
17.0-17.5 &  1509 &   120 &  1123 &    45 &   893 &    40 &   563 &    32 &   425 &    49 \\
17.5-18.0 &  1852 &   113 &  1483 &    46 &  1133 &    40 &   737 &    30 &   584 &    39 \\
18.0-18.5 &  1940 &   113 &  1839 &    46 &  1288 &    39 &   963 &    30 &   748 &    38 \\
18.5-19.0 &  2421 &   145 &  1745 &    50 &  1412 &    40 &  1070 &    30 &   795 &    37 \\
19.0-19.5 &  1843 &   145 &  1503 &    52 &  1198 &    38 &   984 &    30 &   812 &    36 \\
19.5-20.0 &  1629 &   184 &  1396 &    60 &  1257 &    44 &   982 &    30 &   767 &    36 \\
20.0-20.5 &       &       &  1436 &   100 &  1367 &    52 &   918 &    32 &   882 &    37 \\
20.5-21.0 &       &       &  1086 &   396 &  1444 &   111 &  1176 &    43 &   953 &    44 \\
21.0-21.5 &       &       &       &       &  1694 &   309 &  1537 &    80 &  1154 &    65 \\
21.5-22.0 &       &       &       &       &  2154 &   384 &  1849 &   169 &  1259 &   159 \\
22.0-22.5 &       &       &       &       &  1565 &   391 &  1601 &   175 &       &       \\
22.5-23.0 &       &       &       &       &       &       &  1488 &   395 &       &       
\enddata
\end{deluxetable}

\clearpage
\begin{deluxetable}{c|rr|rr|rr|rr|rr|}
\rotate
\tabletypesize{\footnotesize}
\tablecaption{\label{table:mass60} Mass functions for
  60'' annular bins}
\tablehead{
\colhead{Radius}  & \multicolumn{2}{|c|}{60} & \multicolumn{2}{|c|}{120} 
& \multicolumn{2}{|c|}{180} & \multicolumn{2}{|c|}{240} & \multicolumn{2}{|c|}{300} \\
\colhead{Mass}  & \colhead{$\zeta$} & \colhead{$\sigma_{\zeta}$}  
 & \colhead{$\zeta$} & \colhead{$\sigma_{\zeta}$} & \colhead{$\zeta$} & \colhead{$\sigma_{\zeta}$}
 & \colhead{$\zeta$} & \colhead{$\sigma_{\zeta}$} & \colhead{$\zeta$} & \colhead{$\sigma_{\zeta}$}
}
\startdata
0.772-0.811 & 13563 &   347 &  5895 &   208 &  3461 &   853 &       &       &       &       \\
0.724-0.772 & 14407 &   286 &  6559 &   167 &  3556 &   467 &       &       &       &       \\
0.670-0.724 & 14045 &   238 &  7042 &   139 &  3749 &   388 &       &       &       &       \\
0.617-0.670 & 15463 &   274 &  8278 &   141 &  4414 &   392 &       &       &       &       \\
0.567-0.617 & 14128 &   294 &  8261 &   149 &  4763 &   419 &       &       &       &       \\
0.519-0.567 & 14008 &   342 &  8410 &   149 &  4954 &   231 &  3967 &   469 &  2094 &   705 \\
0.467-0.519 & 14140 &   405 &  8424 &   146 &  4998 &   206 &  3645 &   328 &  2176 &   626 \\
0.408-0.467 & 13187 &  1109 &  8416 &   147 &  5264 &   186 &  3868 &   255 &  1780 &   505 \\
0.336-0.408 & 11895 &  1909 &  8556 &   150 &  6120 &   175 &  4365 &   222 &  1951 &   412 \\
0.264-0.336 & 16283 &  2371 &  9655 &   312 &  5882 &   200 &  4873 &   238 &  2052 &   416 \\
0.214-0.264 & 19413 &  3649 & 13728 &  1714 &  7172 &   503 &  4221 &   422 &  2114 &   677 \\
0.176-0.214 &       &       & 16511 &  5445 &  8713 &  2636 &  3625 &   648 &  2010 &  1086 
\enddata
\end{deluxetable}

\clearpage
\begin{deluxetable}{c|rr|rr|rr|rr|rr|}
\rotate
\tabletypesize{\footnotesize}
\tablecaption{\label{table:mass20} Mass functions for
  20'' annular bins}
\tablehead{
\colhead{Radius}  & \multicolumn{2}{|c|}{20} & \multicolumn{2}{|c|}{40} 
& \multicolumn{2}{|c|}{60} & \multicolumn{2}{|c|}{80} & \multicolumn{2}{|c|}{100} \\
\colhead{Mass}  & \colhead{$\zeta$} & \colhead{$\sigma_{\zeta}$}  
 & \colhead{$\zeta$} & \colhead{$\sigma_{\zeta}$} & \colhead{$\zeta$} & \colhead{$\sigma_{\zeta}$}
 & \colhead{$\zeta$} & \colhead{$\sigma_{\zeta}$} & \colhead{$\zeta$} & \colhead{$\sigma_{\zeta}$}
}
\startdata
0.772-0.811 & 19398 &  1539 & 14442 &   584 & 11487 &   514 &  7235 &   406 &  5462 &   624 \\
0.724-0.772 & 19674 &  1196 & 15758 &   488 & 12037 &   420 &  7835 &   322 &  6202 &   412 \\
0.670-0.724 & 17661 &  1025 & 16740 &   419 & 11725 &   352 &  8767 &   275 &  6808 &   348 \\
0.617-0.670 & 22756 &  1363 & 16406 &   469 & 13270 &   379 & 10055 &   285 &  7476 &   351 \\
0.567-0.617 & 18588 &  1465 & 15158 &   526 & 12080 &   387 &  9925 &   302 &  8192 &   367 \\
0.519-0.567 & 16824 &  1903 & 14420 &   620 & 12987 &   456 & 10146 &   306 &  7922 &   367 \\
0.467-0.519 &       &       & 13946 &   973 & 13277 &   509 &  8914 &   310 &  8563 &   362 \\
0.408-0.467 &       &       &       &       & 12388 &   950 & 10091 &   368 &  8174 &   373 \\
0.336-0.408 &       &       &       &       & 11872 &  2168 & 10771 &   563 &  8084 &   453 \\
0.264-0.336 &       &       &       &       & 14840 &  2646 & 12735 &  1161 &  8673 &  1092 \\
0.214-0.264 &       &       &       &       & 15758 &  3940 & 16123 &  1761 &       &       \\
0.176-0.214 &       &       &       &       &       &       & 19507 &  5180 &       &       
\enddata
\end{deluxetable}

\clearpage
\begin{deluxetable}{|l|r|r|r|}
\tabletypesize{\small}
\tablecaption{\label{table:modelparams} Parameters for the King-Michie
  model. References are (1) \citet{Harris1996} (2) \citet{Peterson1994}
}
\tablehead{
\colhead{Parameter} & \colhead{Simulation value} & 
\colhead{Literature value} & \colhead{Reference} }
\startdata
Core radius              & 60'' & 85'' & 1 \\
Concentration            & 1.3  & 1.3  & 1 \\
Tidal radius             & 29'  & 29'  & 1 \\
Velocity dispersion      & 7 km s$^{-1}$ & 7 km s$^{-1}$ & 2 \\
\enddata
\end{deluxetable}


\end{document}